\documentclass[prl,twocolumn,superscriptaddress,showpacs]{revtex4}
\usepackage{times}
\usepackage{amsmath}
\usepackage{amsfonts}
\usepackage{amssymb}
\usepackage{graphicx}
\usepackage{pst-all}

\begin{document}

\title{Splitting the Topological Degeneracy of non-Abelian Anyons}
\author{Parsa Bonderson}
\affiliation{Microsoft Research, Station Q, Elings Hall, University of California,
Santa Barbara, CA 93106}
\date{\today}

\begin{abstract}
We examine tunneling of topological charge between non-Abelian anyons as a perturbation of the long-range effective theory of a topologically ordered system. We obtain energy corrections in terms of the anyons' universal algebraic structure and non-universal tunneling amplitudes. We find that generic tunneling completely lifts the topological degeneracy of non-Abelian anyons. This degeneracy splitting is exponentially suppressed for long distances between anyons, but leaves no topological protection at shorter distances. We also show that general interactions of anyons can be expressed in terms of the transfer of topological charge, and thus can be treated effectively as tunneling interactions.
\end{abstract}


\maketitle

Non-Abelian anyons are quasiparticle excitations in $2+1$ dimensional topologically ordered phases of matter with exotic exchange statistics~\cite{Leinaas77,Goldin85,Fredenhagen89,Froehlich90}. Recently, experimental evidence of such non-Abelian phases has been accumulating, in particular for the $\nu=5/2$ fractional quantum Hall (FQH) state~\cite{Dolev08,Radu08,Willett09a}. The defining properties of non-Abelian anyons are that they can possess a non-local degenerate Hilbert space, even when all of the anyons' local degrees of freedom (i.e. position, spin, etc.) are fixed, and that their exchange acts upon this space via (possibly non-commuting) multi-dimensional representations of the braid group. Remarkably, as long as the anyons are sufficiently separated in space, this topological state space is impervious to local perturbations and the braiding transformations acting upon it are exact. This provides an intrinsic error-protection that makes this state space an ideal place to store and process quantum information~\cite{Kitaev03,Preskill98,Freedman98,Freedman03b,Preskill-lectures}.

However, while these remarkable qualities hold up to corrections exponentially suppressed in the distances between anyons, it is intuitively clear that they must falter below some length scale where microscopic effects become significant and lift the topological state space degeneracy. The microscopic physics of this degeneracy splitting has recently been examined for Ising anyons in the specific context of Kitaev's honeycomb model~\cite{Lahtinen08}, the Moore-Read FQH state~\cite{Baraban09}, and $p_{x}+ip_{y}$ superconductors~\cite{Cheng09}, but a general investigation of the degeneracy splitting of anyons is still absent. In this letter, we examine the degeneracy splitting of arbitrary non-Abelian anyons in a model independent manner by treating the underlying microscopic details as a perturbation within the topological effective theory that tunnels topological charge between anyons. We find that generic tunneling fully lifts the topological degeneracy of non-Abelian anyons. We also show that arbitrary interactions within the topological theory can effectively be treated as tunneling of topological charge.

The long-range effective theory describing quasiparticles in a topologically ordered system is given by an anyon model, which encodes the purely topological properties of anyons, independent of any particular physical representation. We begin with a brief overview of the relevant properties of anyon models (see, e.g. Refs.~\cite{Preskill-lectures,Kitaev06a,Bonderson07b,Bonderson07c} for more details). An anyon model is defined by: a set of conserved quantum numbers called topological charge; fusion rules specifying what can result from combining or splitting topological charges; and braiding rules specifying what happens when the positions of objects carrying topological charge are exchanged. There is a unique ``vacuum'' charge, denoted $I$ or $0$, for with which fusion and braiding is trivial.

Each anyon carries a definite localized value of topological charge. The non-local Hilbert space of a collection of anyons is defined by how the various topological charges can be combined, as dictated by an anyon model's fusion algebra
\begin{equation}
a \times b = \sum_{c} N_{ab}^{c} c
,
\end{equation}
where $N_{ab}^{c}$ are non-negative integers indicating the number of ways topological charges $a$ and $b$ can combine to produce charge $c$. These fusion products and sums can be thought of as similar to tensor products and direct sums of representations in group theory, but without access to the internal degrees of freedom within a representation. We refer to anyons with charges $a$ and $b$ with multiple fusion channels, i.e. $\sum_{c} N_{ab}^{c} >1$, as non-Abelian~\footnote{Strictly speaking, non-Abelian anyons could potentially have Abelian braiding by this definition.}.

It is useful to employ the diagrammatic representation of anyonic states and operators in the effective theory~\footnote{Though these diagrams are designed to resemble worldlines (and in fact the natural diagrammatic representation of an operator corresponding to a spacetime process is exactly its worldlines), we caution the reader that they actually represent operators, and can have multiple equivalent diagrammatic representations, so the diagrams cannot always be interpreted directly as the corresponding worldlines.}. The $N_{ab}^{c}$ different ways that $a$ and $b$ can fuse to give $c$ correspond to orthonormal basis vectors of the fusion/splitting Hilbert spaces $\mathcal{V}_{ab}^{c}$ and $\mathcal{V}^{ab}_{c}$. These vectors are associated with trivalent vertices having labels corresponding to the fusion/splitting:
\begin{eqnarray}
\left( d_{c} / d_{a}d_{b} \right) ^{1/4}
\pspicture[0.6](-0.1,-0.2)(1.5,-1.2)
  \small
  \psset{linewidth=0.9pt,linecolor=black,arrowscale=1.5,arrowinset=0.15}
  \psline{-<}(0.7,0)(0.7,-0.35)
  \psline(0.7,0)(0.7,-0.55)
  \psline(0.7,-0.55) (0.25,-1)
  \psline{-<}(0.7,-0.55)(0.35,-0.9)
  \psline(0.7,-0.55) (1.15,-1)	
  \psline{-<}(0.7,-0.55)(1.05,-0.9)
  \rput[tl]{0}(0.4,0){$c$}
  \rput[br]{0}(1.4,-0.95){$b$}
  \rput[bl]{0}(0,-0.95){$a$}
 \scriptsize
  \rput[bl]{0}(0.85,-0.5){$\mu$}
  \endpspicture
&=&\left\langle a,b;c,\mu \right| \in
\mathcal{V}_{ab}^{c} ,
\label{eq:bra}
\\
\left( d_{c} / d_{a}d_{b}\right) ^{1/4}
\pspicture[0.4643](-0.1,-0.2)(1.5,1.2)
  \small
  \psset{linewidth=0.9pt,linecolor=black,arrowscale=1.5,arrowinset=0.15}
  \psline{->}(0.7,0)(0.7,0.45)
  \psline(0.7,0)(0.7,0.55)
  \psline(0.7,0.55) (0.25,1)
  \psline{->}(0.7,0.55)(0.3,0.95)
  \psline(0.7,0.55) (1.15,1)	
  \psline{->}(0.7,0.55)(1.1,0.95)
  \rput[bl]{0}(0.4,0){$c$}
  \rput[br]{0}(1.4,0.8){$b$}
  \rput[bl]{0}(0,0.8){$a$}
 \scriptsize
  \rput[bl]{0}(0.85,0.35){$\mu$}
  \endpspicture
&=&\left| a,b;c,\mu \right\rangle \in
\mathcal{V}_{c}^{ab},
\label{eq:ket}
\end{eqnarray}
where $\mu = 1, \ldots, N_{ab}^{c}$. The normalization factors involving $d_{a}$, the quantum dimension of the charge $a$, are included so that diagrams are in the isotopy invariant convention. States and operators involving multiple anyons are constructed by appropriately stacking together diagrams, making sure to conserve charge when connecting endpoints of lines. It is clear that the dimension of the topological state space increases as one includes more non-Abelian anyons.

The associativity of fusion is a particularly important property. It is encoded by the unitary (change of basis) isomorphisms
$F^{abc}_{d} : \bigoplus_{e} \mathcal{V}^{ab}_{e} \otimes \mathcal{V}^{ec}_{d} \rightarrow \bigoplus_{e} \mathcal{V}^{af}_{d} \otimes \mathcal{V}^{bc}_{f}$, similar to the $6j$-symbols of angular momentum representations. Diagrammatically, this is represented by
\begin{equation}
\label{eq:F}
\pspicture[0.4545](0,-0.4)(1.8,1.8)
  \small
  \psset{linewidth=0.9pt,linecolor=black,arrowscale=1.5,arrowinset=0.15}
  \psline(1,0.5)(1,0)
  \psline(0.2,1.5)(1,0.5)
  \psline(1.8,1.5) (1,0.5)
  \psline(0.6,1) (1,1.5)
   \psline{->}(0.6,1)(0.3,1.375)
   \psline{->}(0.6,1)(0.9,1.375)
   \psline{->}(1,0.5)(1.7,1.375)
   \psline{->}(1,0.5)(0.7,0.875)
   \psline{->}(1,0)(1,0.375)
   \rput[bl]{0}(0.05,1.6){$a$}
   \rput[bl]{0}(0.95,1.6){$b$}
   \rput[bl]{0}(1.75,1.6){$c$}
   \rput[bl]{0}(0.5,0.5){$e$}
   \rput[bl]{0}(.95,-0.3){$d$}
  \scriptsize
  \rput[br]{0}(0.55,0.85){$\alpha$}
  \rput[br]{0}(1.3,0.4){$\beta$}
  \endpspicture
= \sum_{f,\mu,\nu} \left[F_{d}^{abc}\right]_{\left(e, \alpha, \beta \right) \left(f, \mu, \nu \right)}
\pspicture[0.4545](0,-0.4)(1.8,1.8)
  \small
  \psset{linewidth=0.9pt,linecolor=black,arrowscale=1.5,arrowinset=0.15}
  \psline(1,0.5)(1,0)
  \psline(0.2,1.5)(1,0.5)
  \psline(1.8,1.5) (1,0.5)
  \psline(1.4,1) (1,1.5)
   \psline{->}(0.6,1)(0.3,1.375)
   \psline{->}(1.4,1)(1.1,1.375)
   \psline{->}(1,0.5)(1.7,1.375)
   \psline{->}(1,0.5)(1.3,0.875)
   \psline{->}(1,0)(1,0.375)
   \rput[bl]{0}(0.05,1.6){$a$}
   \rput[bl]{0}(0.95,1.6){$b$}
   \rput[bl]{0}(1.75,1.6){$c$}
   \rput[bl]{0}(1.25,0.45){$f$}
   \rput[bl]{0}(.95,-0.3){$d$}
  \scriptsize
  \rput[br]{0}(1.6,0.85){$\mu$}
  \rput[br]{0}(0.9,0.4){$\nu$}
  \endpspicture
.
\end{equation}
Consistency of these fusion $F$-symbols are ensured by requiring them to satisfy the coherence conditions~\cite{MacLane63} (also referred to as polynomial equations). Indeed, all possible sets of $F$-symbols for a given fusion algebra may be obtained by solving these consistency conditions.

Another important unitary transformation is
\begin{equation}
\label{eq:F_general}
\pspicture[0.4681](0.1,-1.2)(1.7,1.15)
  \small
  \psset{linewidth=0.9pt,linecolor=black,arrowscale=1.5,arrowinset=0.15}
  \psline(0.35,-0.775)(0.35,0.775)
  \psline{>-}(0.35,0.3)(0.35,0.6)
  \psline{<-}(0.35,-0.3)(0.35,-0.6)
  \psline(1.35,-0.775)(1.35,0.775)
  \psline{>-}(1.35,0.3)(1.35,0.6)
  \psline{<-}(1.35,-0.3)(1.35,-0.6)
  \psline(0.35,0.1)(1.35,-0.1)
   \psline{->}(1.35,-0.1)(0.7,0.03)
  \rput[bl]{0}(0.25,0.875){$a$}
  \rput[bl]{0}(1.25,0.875){${b}$}
  \rput[bl]{0}(0.25,-1.075){$c$}
  \rput[bl]{0}(1.25,-1.075){${d}$}
  \rput[br]{0}(0.95,0.13){$e$}
  \scriptsize
  \rput[br]{0}(0.28,0.05){$\alpha$}
  \rput[br]{0}(1.62,-0.25){$\beta$}
\endpspicture
= \sum_{f,\mu,\nu} \left[ F^{ab}_{cd} \right]_{ \left( e,\alpha,\beta \right) \left( f,\mu,\nu \right) }
\pspicture[0.5116](0,-0.85)(1.3,1.3)
 \small
  \psset{linewidth=0.9pt,linecolor=black,arrowscale=1.5,arrowinset=0.15}
  \psline{->}(0.7,0)(0.7,0.45)
  \psline(0.7,0)(0.7,0.55)
  \psline(0.7,0.55) (0.2,1.05)
  \psline{->}(0.7,0.55)(0.3,0.95)
  \psline(0.7,0.55) (1.2,1.05)
  \psline{->}(0.7,0.55)(1.1,0.95)
  \rput[bl]{0}(0.85,0.2){$f$}
  \rput[bl]{0}(1.1,1.15){$b$}
  \rput[bl]{0}(0.1,1.15){$a$}
  \psline(0.7,0) (0.2,-0.5)
  \psline{-<}(0.7,0)(0.35,-0.35)
  \psline(0.7,0) (1.2,-0.5)
  \psline{-<}(0.7,0)(1.05,-0.35)
  \rput[bl]{0}(1.1,-0.8){$d$}
  \rput[bl]{0}(0.1,-0.8){$c$}
  \scriptsize
  \rput[bl]{0}(0.38,0.4){$\mu$}
  \rput[bl]{0}(0.38,-0.0){$\nu$}
  \endpspicture
,
\end{equation}%
which is related to the associativity of Eq.~(\ref{eq:F}) by~\cite{Bonderson07b,Bonderson07c}
\begin{equation}
\left[ F^{ab}_{cd} \right]_{ \left( e,\alpha,\beta \right) \left( f,\mu,\nu \right) }=
\sqrt{ \frac{ d_{e} d_{f} }{ d_{a} d_{d} } } \left[ F^{ceb}_{f} \right]^{\ast}_{ \left( a,\alpha,\mu \right) \left( d,\beta,\nu \right) }
.
\end{equation}

We now begin with an ``unperturbed'' theory which is simply the long-distance effective theory described by the anyon model corresponding to a particular system, i.e. we focus on the (degenerate) ground-states of the system with quasiparticles and ignore states above the gap. When the distance between anyons is large, the effects of the system's microscopic details are weak and can be treated as a perturbation within the effective theory. If the anyons are brought sufficiently close to each other, they will have strong interactions and may even physically fuse~\footnote{Below some separation distance, the notion of being distinct localized objects that carry topological charge no longer applies for a pair of quasiparticles. At this point, the pair are considered to have fused.} making a perturbative treatment inapplicable. We consider the interaction between a pair of anyons, carrying topological charge $a$ and $b$, respectively, in the perturbative regime. Furthermore, we will assume that the interactions do not change the localized topological charges $a$ and/or $b$ of these anyons. This is justified by recognizing that different localized charges have different energetic costs, and assuming that the system has already relaxed into the lowest energy configuration. The leading order interaction is due to simple tunneling of topological charge (virtual anyons) between the anyons. The tunneling charge $e$ must therefore be able to fuse with both $a$ and $b$ without changing them, i.e. $N_{ae}^{a} N_{be}^{b} \neq 0$. There are always such non-trivial tunneling charges $e \neq I$ when $a$ and $b$ have multiple fusion channels, since $\sum_{e} N_{ae}^{a} N_{be}^{b} = \sum_{c} \left( N_{ab}^{c} \right)^{2}$.

To initially provide a less complicated analysis, we begin by considering anyon models with no fusion multiplicities (i.e. $N_{ab}^{c} = 0$ or $1$), which includes all the most physically relevant cases. For such anyons models, we may leave all the vertex labels (greek indices) implicit and assume that diagrams and $F$-symbols with vertices in violation of the fusion rules evaluate to zero. (We also note that such anyons have the same number of fusion channel charges $c$ as tunneling charges $e$.)
The leading order interaction between anyons $a$ and $b$ is given by the tunneling Hamiltonian
\begin{eqnarray}
\label{eq:V1}
&& \!\!\!\!\!\! V_{1} = \sum_{e} \left( \Gamma_{e} \frac{1}{\sqrt{d_{e}}}
\pspicture[0.4681](0.1,-1.2)(1.5,1.15)
  \small
  \psset{linewidth=0.9pt,linecolor=black,arrowscale=1.5,arrowinset=0.15}
  \psline(0.35,-0.775)(0.35,0.775)
  \psline{>-}(0.35,0.3)(0.35,0.6)
  \psline{<-}(0.35,-0.3)(0.35,-0.6)
  \psline(1.35,-0.775)(1.35,0.775)
  \psline{>-}(1.35,0.3)(1.35,0.6)
  \psline{<-}(1.35,-0.3)(1.35,-0.6)
  \psline(1.35,0.1)(0.35,-0.1)
   \psline{->}(0.35,-0.1)(1,0.03)
  \rput[bl]{0}(0.25,0.875){$a$}
  \rput[bl]{0}(1.25,0.875){${b}$}
  \rput[bl]{0}(0.25,-1.075){$a$}
  \rput[bl]{0}(1.25,-1.075){${b}$}
  \rput[br]{0}(0.95,0.13){$e$}
\endpspicture
+\Gamma_{e}^{\ast} \frac{1}{\sqrt{d_{e}}}
\pspicture[0.4681](0.1,-1.2)(1.5,1.15)
  \small
  \psset{linewidth=0.9pt,linecolor=black,arrowscale=1.5,arrowinset=0.15}
  \psline(0.35,-0.775)(0.35,0.775)
  \psline{>-}(0.35,0.3)(0.35,0.6)
  \psline{<-}(0.35,-0.3)(0.35,-0.6)
  \psline(1.35,-0.775)(1.35,0.775)
  \psline{>-}(1.35,0.3)(1.35,0.6)
  \psline{<-}(1.35,-0.3)(1.35,-0.6)
  \psline(0.35,0.1)(1.35,-0.1)
   \psline{->}(1.35,-0.1)(0.7,0.03)
  \rput[bl]{0}(0.25,0.875){$a$}
  \rput[bl]{0}(1.25,0.875){${b}$}
  \rput[bl]{0}(0.25,-1.075){$a$}
  \rput[bl]{0}(1.25,-1.075){${b}$}
  \rput[br]{0}(0.95,0.13){$e$}
\endpspicture
\right) \notag \\
&&= \sum_{e, c} \left( \Gamma_{e} \left[ F^{aeb}_{c} \right]_{ ab} + \Gamma_{e}^{\ast} \left[ F^{aeb}_{c} \right]_{ ab}^{\ast} \right) \sqrt{\frac{d_{c}}{d_{a} d_{b}}}
\,\,
\pspicture[0.5116](0.1,-0.85)(1.3,1.3)
 \small
  \psset{linewidth=0.9pt,linecolor=black,arrowscale=1.5,arrowinset=0.15}
  \psline{->}(0.7,0)(0.7,0.45)
  \psline(0.7,0)(0.7,0.55)
  \psline(0.7,0.55) (0.2,1.05)
  \psline{->}(0.7,0.55)(0.3,0.95)
  \psline(0.7,0.55) (1.2,1.05)
  \psline{->}(0.7,0.55)(1.1,0.95)
  \rput[bl]{0}(0.88,0.2){$c$}
  \rput[bl]{0}(1.1,1.15){$b$}
  \rput[bl]{0}(0.1,1.15){$a$}
  \psline(0.7,0) (0.2,-0.5)
  \psline{-<}(0.7,0)(0.35,-0.35)
  \psline(0.7,0) (1.2,-0.5)
  \psline{-<}(0.7,0)(1.05,-0.35)
  \rput[bl]{0}(1.1,-0.8){$b$}
  \rput[bl]{0}(0.1,-0.8){$a$}
  \endpspicture
\notag \\
&&= \sum_{e,c} \left( \Gamma_{e} \left[ F^{aeb}_{c} \right]_{ ab } + \Gamma_{e}^{\ast} \left[ F^{aeb}_{c} \right]_{ ab }^{\ast} \right) \left| a,b;c \right\rangle \left\langle a,b;c \right|
\end{eqnarray}%
which describes simple tunneling of topological charge between the anyons~\footnote{The second diagram on the first line could be re-written in terms of the first diagram, but it is more convenient to write it this way to make obvious the Hermitian nature of $V_1$.}. This approximation can be improved by adding terms corresponding to processes which decay more quickly as the distance between the anyons increases. The tunneling amplitudes $\Gamma_{e}$ of topological charge $e$ are not universal and depend on the microscopic details of the system in question. Of course, because topological theories have an excitation gap or correlation length, these tunneling amplitudes will generally be exponentially suppressed as $e^{-L/\xi_{e}}$ where $L$ is the distance between the two anyons carrying charges $a$ and $b$, and $\xi_{e}$ is some characteristic length scale for tunneling charge $e$ related to the gap or correlation length. This is akin to the exponential suppression $e^{-mL}$ for tunneling of massive particles~\footnote{For example, a free massive scalar boson in $2+1$ dimensions has the equal time propagator (i.e. two point correlation) $D\left( z_{1}, z_{2} \right)= \frac{-i}{4 \pi \left| z_{1} - z_{2} \right|} e^{-m \left| z_{1} - z_{2} \right| }$.}, and in some cases directly related, such as for $2+1$ dimensional gauge theories, which are topologically massive when there is a Chern-Simons term~\cite{Schonfeld81,Deser82}. The ``tunneling'' of the trivial charge $e=I$ is obviously not an actual tunneling, but we can avoid explicitly excluding it from these expressions by simply letting $\Gamma_{I}=0$.

Hence, the leading correction to the energy of the states described by the different fusion channels $c$ is
\begin{equation}
\label{eq:E_c}
E_{c}^{\left(1\right)} = \sum_{e} \left( \Gamma_{e} \left[ F^{aeb}_{c} \right]_{ ab } + \Gamma^{\ast}_{e} \left[ F^{aeb}_{c} \right]_{ ab }^{\ast} \right)
.
\end{equation}
Notice that the interaction is already diagonal in $c$, resulting from the fact that no other anyons are involved. The quantity $\left[ F^{aeb}_{c} \right]_{ ab }$ characterizes the difference in effect on state $\left| a,b;c \right\rangle$ that results from a charge $e$ fusing with $a$ as compared to fusing with $b$. Here, it tells us whether the transfer of topological charge $e$ between $a$ and $b$ can distinguish their different fusion channels $c$. Since $\left[ F_{ab}^{ab} \right]_{ ec }$ is unitary, the matrix
\begin{equation}
T_{ ec } = \left[ F^{aeb}_{c} \right]_{ ab } = \sqrt{ \frac{ d_{a} d_{b} }{ d_{c} d_{e} } } \left[ F_{ab}^{ab} \right]^{\ast}_{ ec }
\end{equation}
can be inverted to give
\begin{equation}
T^{-1}_{ ce } = \frac{ d_{c} d_{e} }{ d_{a} d_{b} } \left[ F^{aeb}_{c} \right]^{\ast}_{ ab }
.
\end{equation}
This implies that for generic values of the tunneling coefficients $\Gamma_{e}$, the shifts in energy $E_{c}^{\left(1\right)}$ will be different for all $c$. In other words, the degeneracy of fusion channels $c$ of $a$ and $b$ will generically be completely lifted.

It is now straightforward to return to the general case where fusion multiplicities are allowed. If we reconstitute the vertex labels for the diagrams of Eq.~(\ref{eq:V1}) and use the corresponding tunneling amplitudes $\Gamma_{e, \alpha, \beta}$, the energy corrections are now obtained by diagonalizing the $N_{ab}^{c}$ by $N_{ab}^{c}$ Hermitian matrices (with indices $\mu$ and $\nu$)
\begin{eqnarray}
V_{c,\mu \nu}^{\left(1\right)} &=& \sum_{e,\alpha,\beta} \left( \Gamma_{e, \alpha, \beta} \left[ F^{aeb}_{c} \right]_{ \left( a,\alpha,\nu \right) \left( b,\beta,\mu \right) } \right. \notag \\
&& \qquad \qquad \left. + \Gamma^{\ast}_{e, \alpha, \beta} \left[ F^{aeb}_{c} \right]_{ \left( a,\alpha,\mu \right) \left( b,\beta,\nu \right) }^{\ast} \right)
\end{eqnarray}
corresponding to the charge $c$ fusion channels of $a$ and $b$ (the perturbation is already diagonal in $c$). We now have
\begin{eqnarray}
T_{ \left( e,\alpha,\beta \right) \left( c,\mu,\nu \right) } &=& \left[ F^{aeb}_{c} \right]_{ \left( a,\alpha,\nu \right) \left( b,\beta,\mu \right) } \\
T^{-1}_{ \left( c,\mu,\nu \right) \left( e,\alpha,\beta \right) } &=& \frac{ d_{c} d_{e} }{ d_{a} d_{b} } \left[ F^{aeb}_{c} \right]^{\ast}_{ \left( a,\alpha,\nu \right) \left( b,\beta,\mu \right) }
.
\end{eqnarray}
Again we see that for generic values of the tunneling amplitudes $\Gamma_{e, \alpha, \beta}$, the energy degeneracy will be completely lifted. However, one might expect that in some cases the tunneling amplitudes will obey a symmetry, for example $\Gamma_{e, \alpha, \beta} = \Gamma_{e}$ if tunneling is independent of the particular fusion channels $\alpha$ and $\beta$ involved. It is difficult to predict in generality whether such symmetries will occur and what effect they will have on the internal degeneracy within a fusion space $\mathcal{V}_{c}^{ab}$; however, one still generically finds splitting of the energies for different $c$. (See Eq.~\ref{eq:V2} for an interaction that generically lifts the degeneracy for different $c$, but leaves the spaces $\mathcal{V}_{c}^{ab}$ degenerate.) Even if degeneracy within subspaces $\mathcal{V}_{c}^{ab}$ remains, utilizing these protected subspaces for quantum information processing would likely be impractical (if not impossible), because braiding transformations and methods of distinguishing states within the subspaces are significantly more limited.

The above analysis of the tunneling perturbation $V_{1}$ only appealed to the fusion properties of anyons, so in principle it could also apply to any system described by a unitary fusion category, whether or not it also has braiding.

There are also braiding processes associated with lifting fusion channel degeneracies, such as when anyons pair-created from vacuum braid around both anyons $a$ and $b$ and then re-annihilate into vacuum. This is described by
\begin{eqnarray}
\label{eq:V2}
V_{2} &=& \sum_{z} \left( \gamma_{z} \frac{1}{d_{z}}
\pspicture[0.4681](0.2,-1.2)(1.6,1.15)
  \small
  \psset{linewidth=0.9pt,linecolor=black,arrowscale=1.5,arrowinset=0.15}
  \psline(0.55,-0.775)(0.55,0.775)
  \psline(1.15,-0.775)(1.15,0.775)
  \psellipse[linewidth=0.9pt,linecolor=black,border=.08](0.85,0)(0.5,0.3)
  \psline[linewidth=0.9pt,linecolor=black,arrows=->,arrowscale=1.2,arrowinset=0.15](1.354,-0.01)(1.364,0.01)
  \psline[border=2.5pt](0.55,0.1)(0.55,0.4)
  \psline[arrows=>-](0.55,0.3)(0.55,0.6)
  \psline[border=2.5pt](1.15,0.1)(1.15,0.4)
  \psline[arrows=>-](1.15,0.3)(1.15,0.6)
  \rput[bl]{0}(0.45,0.875){$a$}
  \rput[bl]{0}(1.05,0.875){${b}$}
  \rput[bl]{0}(0.45,-1.075){$a$}
  \rput[bl]{0}(1.05,-1.075){${b}$}
  \rput[bl]{0}(1.4,-0.3){${z}$}
\endpspicture
+
\gamma_{z}^{\ast} \frac{1}{d_{z}}
\pspicture[0.4681](0.2,-1.2)(1.6,1.15)
  \small
  \psset{linewidth=0.9pt,linecolor=black,arrowscale=1.5,arrowinset=0.15}
  \psline(0.55,-0.775)(0.55,0.775)
  \psline(1.15,-0.775)(1.15,0.775)
  \psellipse[linewidth=0.9pt,linecolor=black,border=.08](0.85,0)(0.5,0.3)
  \psline[linewidth=0.9pt,linecolor=black,arrows=<-,arrowscale=1.2,arrowinset=0.15](1.214,-0.218)(1.23,-0.20)
  \psline[border=2.5pt](0.55,0.1)(0.55,0.4)
  \psline[arrows=>-](0.55,0.3)(0.55,0.6)
  \psline[border=2.5pt](1.15,0.1)(1.15,0.4)
  \psline[arrows=>-](1.15,0.3)(1.15,0.6)
  \rput[bl]{0}(0.45,0.875){$a$}
  \rput[bl]{0}(1.05,0.875){${b}$}
  \rput[bl]{0}(0.45,-1.075){$a$}
  \rput[bl]{0}(1.05,-1.075){${b}$}
  \rput[bl]{0}(1.4,-0.3){${z}$}
\endpspicture
\right)
\notag \\
&=& \sum_{z,c,\mu} \left( \gamma_{z} M_{zc} + \gamma^{\ast}_{z} M^{\ast}_{zc} \right) \sqrt{\frac{d_{c}}{d_{a} d_{b}}}
\pspicture[0.5116](0,-0.85)(1.3,1.3)
 \small
  \psset{linewidth=0.9pt,linecolor=black,arrowscale=1.5,arrowinset=0.15}
  \psline{->}(0.7,0)(0.7,0.45)
  \psline(0.7,0)(0.7,0.55)
  \psline(0.7,0.55) (0.2,1.05)
  \psline{->}(0.7,0.55)(0.3,0.95)
  \psline(0.7,0.55) (1.2,1.05)
  \psline{->}(0.7,0.55)(1.1,0.95)
  \rput[bl]{0}(0.85,0.2){$c$}
  \rput[bl]{0}(1.1,1.15){$b$}
  \rput[bl]{0}(0.1,1.15){$a$}
  \psline(0.7,0) (0.2,-0.5)
  \psline{-<}(0.7,0)(0.35,-0.35)
  \psline(0.7,0) (1.2,-0.5)
  \psline{-<}(0.7,0)(1.05,-0.35)
  \rput[bl]{0}(1.1,-0.8){$b$}
  \rput[bl]{0}(0.1,-0.8){$a$}
  \scriptsize
  \rput[bl]{0}(0.38,0.4){$\mu$}
  \rput[bl]{0}(0.38,-0.0){$\mu$}
  \endpspicture
\notag \\
&=& \sum_{z,c,\mu } \left( \gamma_{z} M_{zc} + \gamma^{\ast}_{z} M^{\ast}_{zc} \right) \left| a,b;c,\mu \right\rangle \left\langle a,b;c,\mu \right|
.
\end{eqnarray}%
Clearly, this will be a smaller perturbation than $V_1$ (for $L$ large), since the distance the virtual anyon must travel is about twice that for the tunneling case, and hence the amplitude for this process $\left| \gamma_{z} \right| \sim e^{-2L/\xi_{z}} \sim \left| \Gamma_{z} \right|^{2} $. Thus, one only really needs to consider this perturbation when higher order terms are significant, however we will see that this perturbation can in fact be absorbed into $V_{1}$. The resulting change in energy from this perturbation is
\begin{equation}
E_{c}^{\left(2\right)} = \sum_{z} \left( \gamma_{z} M_{zc} + \gamma^{\ast}_{z} M^{\ast}_{zc} \right)
,
\end{equation}
where $M_{ab} = \frac{ S_{ab} S_{II} }{ S_{Ia} S_{Ib} }$ is the monodromy scalar component (related to the topological $S$-matrix) which plays a significant role in interference experiments~\cite{Bonderson06b,Bonderson07b,Bonderson07c}. In this context, $M_{zc}$ tells us whether monodromy of charge $z$ can distinguish between different fusion channels $c$~\footnote{We note that $M_{ab}$ is invertible iff $S_{ab}$ is unitary iff the theory is modular (i.e. corresponds to a TQFT), in which case the inverse is $M^{-1}_{ab} = \frac{ S_{Ia} S_{Ib} S_{ab}^{\ast} }{ S_{II} }$.}.

Comparing the forms of these Hamiltonians, we see that the process in $V_{2}$ can be treated effectively as a tunneling of topological charge from $a$ to $b$, and thus absorbed into $V_{1}$ with a redefinition of the tunneling amplitudes. In particular, one could re-write the diagrams in Eq.~(\ref{eq:V2}) in terms of those in Eq.~(\ref{eq:V1}) using the diagrammatic rules. This is just the observation that braiding can have the effect of transferring topological charge between non-Abelian anyons without them ever actually coming into direct contact.

In fact, a bit more thought reveals that the diagrams representing any process in the effective theory can generally be re-written in terms of diagrams representing tunneling processes. Specifically, a completely general interaction $V$ of the topological charges $a$ and $b$ (that leaves the localized charges unchanged) can be represented by
\begin{equation}
\label{eq:V}
V =
\pspicture[0.4681](-0.1,-1.2)(1.8,1.15)
  \small
  \psset{linewidth=0.9pt,linecolor=black,arrowscale=1.5,arrowinset=0.15}
  \psline(0.35,-0.775)(0.35,-0.3)
  \psline(0.35,0.3)(0.35,0.775)
  \psline{>-}(0.35,0.4)(0.35,0.6)
  \psline{<-}(0.35,-0.4)(0.35,-0.6)
  \psline(1.35,-0.775)(1.35,-0.3)
  \psline(1.35,0.3)(1.35,0.775)
  \psline{>-}(1.35,0.4)(1.35,0.6)
  \psline{<-}(1.35,-0.4)(1.35,-0.6)
  \rput[bl]{0}(0.25,0.875){$a$}
  \rput[bl]{0}(1.25,0.875){${b}$}
  \rput[bl]{0}(0.25,-1.075){$a$}
  \rput[bl]{0}(1.25,-1.075){${b}$}
  \psframe[linewidth=0.9pt,linecolor=black,border=0](0.15,-0.3)(1.55,0.3)
  \rput[bl]{0}(0.7,-0.1){$V$}
\endpspicture
= \sum_{c,\mu,\nu} V_{ c,\mu,\nu }
\pspicture[0.5116](0,-0.85)(1.3,1.3)
 \small
  \psset{linewidth=0.9pt,linecolor=black,arrowscale=1.5,arrowinset=0.15}
  \psline{->}(0.7,0)(0.7,0.45)
  \psline(0.7,0)(0.7,0.55)
  \psline(0.7,0.55) (0.2,1.05)
  \psline{->}(0.7,0.55)(0.3,0.95)
  \psline(0.7,0.55) (1.2,1.05)
  \psline{->}(0.7,0.55)(1.1,0.95)
  \rput[bl]{0}(0.85,0.2){$c$}
  \rput[bl]{0}(1.1,1.15){$b$}
  \rput[bl]{0}(0.1,1.15){$a$}
  \psline(0.7,0) (0.2,-0.5)
  \psline{-<}(0.7,0)(0.35,-0.35)
  \psline(0.7,0) (1.2,-0.5)
  \psline{-<}(0.7,0)(1.05,-0.35)
  \rput[bl]{0}(1.1,-0.8){$b$}
  \rput[bl]{0}(0.1,-0.8){$a$}
  \scriptsize
  \rput[bl]{0}(0.38,0.4){$\mu$}
  \rput[bl]{0}(0.38,-0.0){$\nu$}
  \endpspicture
,
\end{equation}%
where $V_{ c,\mu,\nu } = V^{\ast}_{ c,\nu,\mu }$ because $V$ is Hermitian. This can be treated effectively by including it in the tunneling interaction $V_1$ with the addition of the effective amplitudes
\begin{equation}
\label{eq:Gamma_V}
\Gamma_{e, \alpha, \beta}^{\text{eff}} = \frac{1}{2} \sum_{c,\mu,\nu} V_{c,\mu,\nu} T^{-1}_{\left(c, \mu, \nu \right) \left( e, \alpha, \beta \right)}
.
\end{equation}
For example, the effective amplitude from $V_2$ would be
\begin{equation}
\label{eq:Gamma_2}
\Gamma_{e, \alpha, \beta}^{(2)} = \sum_{z,c,\mu} \gamma_{z} M_{zc} T^{-1}_{\left(c, \mu, \mu \right) \left( e, \alpha, \beta \right)}
.
\end{equation}
Thus, even when higher order processes are significant, all interactions between $a$ and $b$ that leave the localized charges unchanged can be represented using only $V_1$ with effective tunneling amplitudes that account for all the different processes. In this way, $\Gamma_{e, \alpha, \beta}$ can be treated as a (non-universal) phenomenological parameter, which one may (attempt to) compute for any particular model, to any desired order.

Similarly, one can show that all interactions of $n$ anyons can be written in terms of the $(n-1)$th order tunneling processes represented by diagrams with a tunneling charge line connecting each adjacent pair of anyons' lines. This makes explicit the fact that the fundamental mechanism which mediates interactions in the long-range effective theory and splits topological degeneracies is the transfer of topological charge between anyons. The type of analysis performed in this letter can be used to guide the modeling of interactions employed in many-body studies of interacting non-Abelian anyons~\cite{Read00,Kitaev06unpub,Grosfeld06a,Feiguin07,Bonesteel07,Gils09}.

We now consider examples of non-Abelian anyon models that are particularly relevant for physical systems:

{\bf Ising} anyons occur in several FQH states likely to exist in the second Landau level~\cite{Moore91,Bonderson07d}, $p_x +ip_y$ superconductors~\cite{Read00}, and Kitaev's honeycomb model~\cite{Kitaev06a}. A pair of $a=b=\sigma$ anyons have fusion channels $c=I,\psi$ and tunneling charges $e=I,\psi$, and
\begin{equation}
\left[ F^{\sigma e \sigma}_{c} \right]_{ \sigma \sigma } = \left[
\begin{array}{rr}
1 & 1 \\
1 & -1
\end{array}
\right]_{ec}
,
\end{equation}
which gives the energy corrections 
\begin{equation}
E_{I}^{\left(1\right)} =-E_{\psi }^{\left(1\right)}= \Gamma_{\psi} + \Gamma_{\psi}^{\ast}
.
\end{equation}
We also note that $\Gamma_{\psi}^{(2)} = \gamma_{\sigma}$ in Eq.~(\ref{eq:Gamma_2}).

Analyses of Ising anyons in Kitaev's honeycomb model~\cite{Lahtinen08}, $p_{x} +ip_{y}$ superconductors~\cite{Cheng09}, and the Moore-Read state~\cite{Baraban09} have all found that while this energy splitting decays exponentially, it also oscillates between positive and negative values as a result of the short-wavelength physics. For the honeycomb model and $p_{x} +ip_{y}$ superconductors, it is known that $E_{I} < E_{\psi} $ for small $L$, since $I$ actually corresponds to no excitations in these cases. For the Moore-Read state, however, it was found that $E_{\psi} < E_{I} $ for small $L$~\cite{Baraban09}.

{\bf Fibonacci} anyons occur in a FQH state that may also exist in the second Landau level~\cite{Read99}. A pair of $a=b=\varepsilon$ anyons have fusion channels $c=I,\varepsilon$ and tunneling charges $e=I,\varepsilon$, and
\begin{equation}
\left[ F^{\varepsilon e \varepsilon}_{c} \right]_{ \varepsilon \varepsilon} = \left[
\begin{array}{cc}
1 & 1 \\
1 & -\phi^{-1}
\end{array}
\right]_{ec}
,
\end{equation}
where $\phi = \frac{1+\sqrt{5}}{2}$ is the Golden ratio, which gives the energy corrections 
\begin{equation}
E_{I}^{\left(1\right)} = \Gamma_{\varepsilon} + \Gamma_{\varepsilon}^{\ast} , \quad E_{\varepsilon }^{\left(1\right)} = -\phi^{-1}\left( \Gamma_{\varepsilon} + \Gamma_{\varepsilon}^{\ast} \right)
.
\end{equation}
We emphasize that this energy splitting is not symmetric.

{\bf SU$(2)_{k}$} anyons are the prototypic examples of non-Abelian anyons~\cite{Witten89}. A pair of $a=b=\frac{1}{2}$ anyons have fusion channels $c=0,1$ and tunneling charges $e=0,1$, and
\begin{equation}
\left[ F^{\frac{1}{2} e \frac{1}{2}}_{c} \right]_{ \frac{1}{2} \frac{1}{2} } = \left[
\begin{array}{cc}
1 & 1 \\
1 & -d_{1}^{-1}
\end{array}
\right]_{ec}
,
\end{equation}
where $d_{1} = 4 \cos^{2} \left( \frac{ \pi }{k+2} \right) -1$ is the quantum dimension of the topological charge $1$, which gives
\begin{equation}
E_{0}^{\left(1\right)} = \Gamma_{1} + \Gamma_{1}^{\ast}, \quad E_{1}^{\left(1\right)} = -d_{1}^{-1}\left( \Gamma_{1} + \Gamma_{1}^{\ast} \right)
.
\end{equation}
For an example with more than two fusion channels, we can consider SU$(2)_{4}$ for a pair of $a=b=1$ anyons. These have fusion channels $c=0,1,2$ and tunneling charges $e=0,1,2$, and
\begin{equation}
\left[ F^{1 e 1}_{c} \right]_{ 11} = \left[
\begin{array}{rrr}
1 & 1  & 1\\
1 & 0  & -1 \\
1 & -1 &  1
\end{array}
\right]_{ec}
,
\end{equation}
which gives
\begin{eqnarray}
E_{0}^{\left(1\right)} &=&  \left( \Gamma_{1} + \Gamma_{1}^{\ast} \right) + \left( \Gamma_{2} + \Gamma_{2}^{\ast} \right), \\
E_{1}^{\left(1\right)} &=& -\left( \Gamma_{2} + \Gamma_{2}^{\ast} \right) \\
E_{2}^{\left(1\right)} &=& -\left( \Gamma_{1} + \Gamma_{1}^{\ast} \right) + \left( \Gamma_{2} + \Gamma_{2}^{\ast} \right)
.
\end{eqnarray}
We notice that if there were only tunneling of the $e=2$ topological charge, the energies of the fusion channels $c=0$ and $2$ would not split.

We have examined topological charge tunneling interactions between anyons, representing perturbations of the long-range effective theory resulting from the microscopic details within the system. We found that these interactions, which become significant as anyons approach each other, will generically completely split the fusion channel degeneracy of non-Abelian anyons. In principle, this energy splitting could be used to perform topological charge measurements, and even to implement computational gates~\cite{Bravyi06,Bonderson08a,Bonderson08b}. However, in practice, the energy splitting will likely be a difficult resource to utilize with sufficient precision, and, even worse, allows the environment to easily couple to the non-local state space. Indeed, if the interactions described here were mediated by real anyons, e.g. produced by thermal or noise perturbations, rather than virtual anyons, then our analysis carries over to show this enables the environment to couple to all the fusion channels and cause decoherence in any quantum information encoded in the topological Hilbert space. Similar to the effects of separation distances, if temperature and noise frequencies are kept small (compared to the gap and correlation scales), then their effects will be exponentially suppressed, but if they are sufficiently large, then their effects will become strong, making the long-distance effective theory inapplicable. As errors in topological quantum information are due to undesired transfer of topological charge, which we have shown leaves no protected subspaces, this letter reaffirms the absolute importance of keeping anyons well-separated and of ensuring that the temperature and noise frequencies in the system are much smaller than the gap in order to capitalize on the topological protection of encoded quantum information.

\begin{acknowledgments}
I would like to thank N.~Bonesteel, M.~Freedman, E.~Grosfeld, A.~Ludwig, J.~Pachos, C.~Nayak, K.~Shtengel, J.~Slingerland, S.~Trebst, and Z.~Wang for illuminating discussions.
\end{acknowledgments}


\end{document}